\documentclass{pasj00}
\draft

\begin{document}
\SetRunningHead{Abe et al.}{Spectroscopy of HAYABUSA Re-entry}
\Received{2011/06/23}%{yyyy/mm/dd}
\Accepted{2011/08/28}%{yyyy/mm/dd}
%\Published{}%{yyyy/mm/dd}

\title{Near-Ultraviolet and Visible Spectroscopy of HAYABUSA Spacecraft Re-entry}

\author{Shinsuke \textsc{Abe}}
\affil{National Central University, 300 Jhongda Road, Jhongli, Taoyuan 32001, Taiwan}\email{shinsuke.avell@gmail.com}
\author{Kazuhisa \textsc{Fujita}}
\affil{Japan Aerospace Exploration Agency, 7-44-1 Jinidaiji-higashi-machi, Chofu, Tokyo 182-8522, Japan}\email{fujita.kazuhisa@jaxa.jp}
\author{Yoshihiro \textsc{Kakinami}}
\affil{Institute of Seismology and Volcanology, Hokkaido University, Kita 10 Nishi 8, Kita-ku, Sapporo 060-0810, Japan}\email{kakinami@mail.sci.hokudai.ac.jp}
\author{Ohmi \textsc{Iiyama}}
\affil{Osaka Science Museum, 4-2-1 Nakanoshima, Kitaku, Osaka, Osaka 530-0005, Japan}
\author{Hirohisa \textsc{Kurosaki}}
\affil{Japan Aerospace Exploration Agency, 7-44-1 Jinidaiji-higashi-machi, Chofu, Tokyo 182-8522, Japan}
\author{Michael A. \textsc{Shoemaker}}
\affil{Kyushu University, 744 Motooka, Nishi-ku, Fukuoka 819-0395, Japan}
\author{Yasuo \textsc{Shiba} and Masayoshi \textsc{Ueda}}
\affil{Nippon Meteor Society, 43-2 Asuka Habikino Osaka 583-0842, Japan}
\and
\author{Masaharu \textsc{Suzuki}}
\affil{GOTO Inc., 4-16 Yazaki-cho, Fuchu, Tokyo 183-8530, Japan}

\KeyWords{Meteors, meteoroids -- asteroid -- atmospheric re-entry -- spectroscopy} %Do NOT move this preamble from here!

\maketitle

\begin{abstract}

HAYABUSA is the first spacecraft ever to land on and lift off from any celestial body other than the moon. The mission, which returned asteroid samples to the Earth while overcoming various technical hurdles, ended on June 13, 2010, with the planned atmospheric re-entry. In order to safely deliver the sample return capsule, the HAYABUSA spacecraft ended its 7-year journey in a brilliant ``artificial fireball'' over the Australian desert. Spectroscopic observation was carried out in the near-ultraviolet and visible wavelengths between 3000 and 7500 \AA~ at 3 -- 20 \AA~ resolution. Approximately 100 atomic lines such as Fe\emissiontype{I}, Mg\emissiontype{I}, Na\emissiontype{I}, Al\emissiontype{I}, Cr\emissiontype{I}, Mn\emissiontype{I}, Ni\emissiontype{I}, Ti\emissiontype{I}, Li\emissiontype{I}, Zn\emissiontype{I}, O\emissiontype{I}, and N\emissiontype{I} were identified from the spacecraft. Exotic atoms such as Cu\emissiontype{I}, Mo\emissiontype{I}, Xe\emissiontype{I} and Hg\emissiontype{I} were also detected. A strong Li\emissiontype{I} line (6708 \AA) at a height of $\sim$55 km originated from the onboard Li-Ion batteries. The FeO molecule bands at a height of $\sim$63 km were probably formed in the wake of the spacecraft. The effective excitation temperature as determined from the atomic lines varied from 4500 K to 6000 K. The observed number density of Fe\emissiontype{I} was about 10 times more abundant than Mg\emissiontype{I} after the spacecraft explosion. N$_{2}^{+}$ ($1^-$) bands from a shock layer and CN violet bands from the sample return capsule's ablating heat shield were dominant molecular bands in the near-ultraviolet region of 3000 -- 4000 \AA. OH($A$-$X$) band was likely to exist around 3092 \AA. A strong shock layer from the HAYABUSA spacecraft was rapidly formed at heights between 93 km and 83 km, which was confirmed by detection of N$_{2}^{+}$ ($1^-$) bands with a vibration temperature of $\sim$13000 K. Gray-body temperature of the capsule at a height of $\sim$42 km was estimated to be $\sim$2437 K which is matched to a theoretical prediction. The final message of the HAYABUSA spacecraft and its sample return capsule are discussed through our spectroscopy.

\end{abstract}

\section{Introduction}

HAYABUSA, the third engineering space mission of JAXA/ISAS (Japan Aerospace eXploration Agency/Institute of Space and Astronautical Sciences) had several engineering technologies to verify in space (e.g, \cite{kawaguchi08}). HAYABUSA was launched on May 9, 2003. On September 12, 2005, HAYABUSA arrived at the asteroid (25143) Itokawa. The mission was the first to reveal that Itokawa is a rubble-pile body rather than a single monolithic asteroid among $S$-class asteroids (\cite{abe06}; \cite{fujiwara06}). Finally, the spacecraft performed a landing on Itokawa to collect asteroid samples in November, 2005. Due to a loss of communications, HAYABUSA started an orbit transfer to return to the Earth in April, 2007. The round-trip travel between the Earth and Itokawa with the aid of ion engine propulsion was the first success of its kind in the world (\cite{kuninaka08}).

On June 13, 2010, the HAYABUSA spacecraft returned to the Earth with the re-entry capsule containing asteroid samples. The sample return capsule was separated at 10:51 UT, which was just 3 hours before the atmospheric re-entry. Due to the failure of all bi-propellant thrusters for orbital maneuvering, the HAYABUSA spacecraft could not escape from its Earth collision course. Though Japan had several successful experiences with re-entry capsule tests (e.g., RFT-2 (1992), OREX (1994), EXPRESS (1995), HYFREX (1996), and USERS (2003)), the HAYABUSA sample return capsule was the first Japanese re-entry opportunity that entered the Earth's atmosphere directly from an interplanetary transfer orbit with a velocity over 12 km/s. The HAYABUSA sample return capsule and the spacecraft entered the atmosphere at 13:51 UT. The HAYABUSA spacecraft disintegrated in the atmosphere, and the capsule flew nominally and landed approximately 500 m from its targeted landing point.

Spectroscopy of the HAYABUSA re-entry was a golden opportunity to understand (i) the atmospheric influence upon Earth impactors such as meteors, meteorites and meter-sized mini-asteroids, because for such natural bodies the original material, mass, and shape are all unknown (e.g., \cite{ceplecha96}; \cite{abe09}); for HAYABUSA, not only these parameters but also the re-entry trajectory were under perfect control, (ii) the hypervelocity impact of large objects that are difficult to reproduce in laboratory experiments, and (iii) the flight environment of re-entry capsules for the utilization of future Japanese sample return missions.

\section{Observation and data reduction}

\begin{table*}
  \caption{Specifications of spectroscopic and imaging cameras.}\label{tb_instrument}
  \begin{center}
    \begin{tabular}{cccccccc}
      \hline
Name        & Color & Lens         & Spectrograph & FOV (H $\times$ V)             & Pixels              & Frame rate \\ \hline
VIS-HDTV    & Color & 80mm/f5.6    & 300 /mm\footnotemark[$\dagger$]  & 25$^{\circ} \times$ 17$^{\circ}$  & 1920 $\times$ 1080 & 30 fps \\
UV-II       & B/W   & 30mm/f1.2    & 600 /mm\footnotemark[$\ddagger$] & 23$^{\circ} \times$ 13$^{\circ}$  & 720 $\times$ 480   & 29.97 (NTSC) \\
NCR-550a\footnotemark[$*$]
            & Color & 4.6 -- 60 mm & --           & 90$^{\circ}$ -- 8$^{\circ}$ diagonal & 800 $\times$ 412 & 29.97 (NTSC) \\
Nikon D300s\footnotemark[$*$]
            & Color & 18 -- 200mm  & --           & 10$^{\circ} \times$ 7$^{\circ}$    & 1280 $\times$ 720 & 24 fps \\
WAT-100N\footnotemark[$**$]
            & B/W   & 25mm/f0.95   & --           & 15$^{\circ} \times$ 11$^{\circ}$   & 768 $\times$ 494  & 29.97 (NTSC) \\
      \hline
    \end{tabular}
  \end{center}
\footnotemark[$\dagger$] Spectrum resolution = 12\AA~ (1st order), 6\AA~ (2nd order) \& 3\AA~ (3rd order).\\
\footnotemark[$\ddagger$] Spectrum resolution = 20\AA.\\
\footnotemark[$*$] Video data was published on the JAXA digital archive; http://jda.jaxa.jp \\
\footnotemark[$**$] Video data is used for the trajectory estimation of fragmented HAYABUSA (\cite{shoemaker11}).
\end{table*}

\begin{table*}
  \caption{Beginning and terminating of light detected by each camera and corresponding trajectory of HAYABUSA and the capsule.}\label{tb_event}
  \begin{center}
    \begin{tabular}{crrcc}
  \hline
Time       & Height\footnotemark[$\dagger$] & Velocity\footnotemark[$\ddagger$] & Event      & Cameras  \\
UT         & \multicolumn{1}{c}{$km$}    & \multicolumn{1}{c}{$km s^{-1}$}    &          & (SPectrum or IMage) \\   \hline
13:51:50.1 & 113.21 & 12.05 & Beginning  & WAT-100N (IM) \\
13:51:53.5 & 106.17 & 12.06 & Beginning  & NCR-550a (IM) \\
13:51:54.3 & 104.53 & 12.06 & Beginning  & UV-II (SP)    \\
13:51:57.8 &  97.46 & 12.06 & Beginning  & D300s (IM)    \\
13:52:03.9 &  85.47 & 12.07 & Beginning  & VIS-HDTV (SP) \\
13:52:12.9 &  68.63 & 11.95 & Explosion  & all \\
13:52:13.6 &  67.38 & 11.90 & Explosion  & all \\
13:52:13.9 &  66.85 & 11.88 & Maximum    & all \\
13:52:17.1 &  61.39 & 11.57 & Explosion  & all \\
13:52:17.2 &  61.22 & 11.55 & Explosion  & all \\
13:52:19.5 &  57.53 & 11.18 & Explosion  & all \\
13:52:19.9 &  56.90 & 11.10 & Explosion  & all \\
13:52:22.7 &  52.79 & 10.44 & Ending     & VIS-HDTV (SP) \\
13:52:40.2 &  36.81 &  3.38 & Ending     & UV-II (SP)    \\
13:52:42.1 &  35.98 &  2.94 & Ending     & D300s (IM)    \\
13:52:47.4 &  34.09 &  2.08 & Ending     & NCR550a (IM)  \\
13:52:50.6 &  33.17 &  1.74 & Ending     & WAT-100N (IM) \\
  \hline
    \end{tabular}
  \end{center}
\footnotemark[$\dagger$] The trajectory of HAYABUSA and the capsule was obtained by our observation (\cite{borovicka11}).\\
\footnotemark[$\ddagger$] The velocity relative to the Earth's center is given (the velocity relative to the surface is about 0.37 km/s lower because of Earth's rotation).
\end{table*}

\begin{figure}
  \begin{center}
%    \FigureFile(85mm,85mm){figure1.eps}
    \FigureFile(\linewidth,1mm){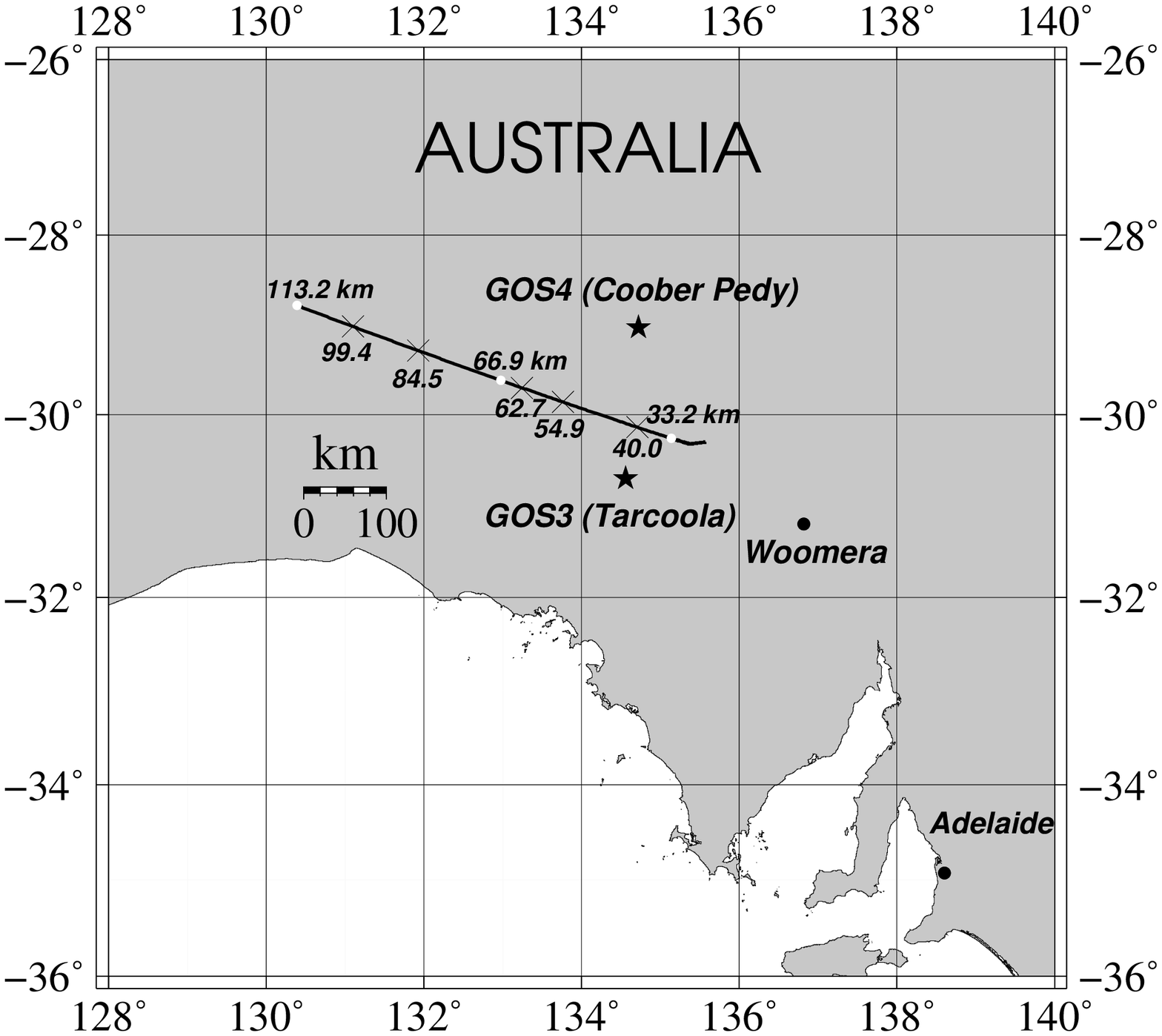}
\vspace{-6cm}
  \end{center}
  \caption{HAYABUSA trajectory and observation stations in Australia. The HAYABUSA re-entry has flown over and landed on the Woomera Prohibited Area (WPA) where unauthorized people have no admittance. GOS sites were located on the border of the WPA. A thick black line shows the trajectory of sample return capsule in which white dots indicate the heights of beginning (113.2 km), maximum brightness (66.9 km) and the end (33.2 km) detected by our cameras. Several heights at which the spectra were analyzed are indicated by the X marks.}\label{map}
\end{figure}

\begin{figure}
  \begin{center}
    \FigureFile(\linewidth,1mm){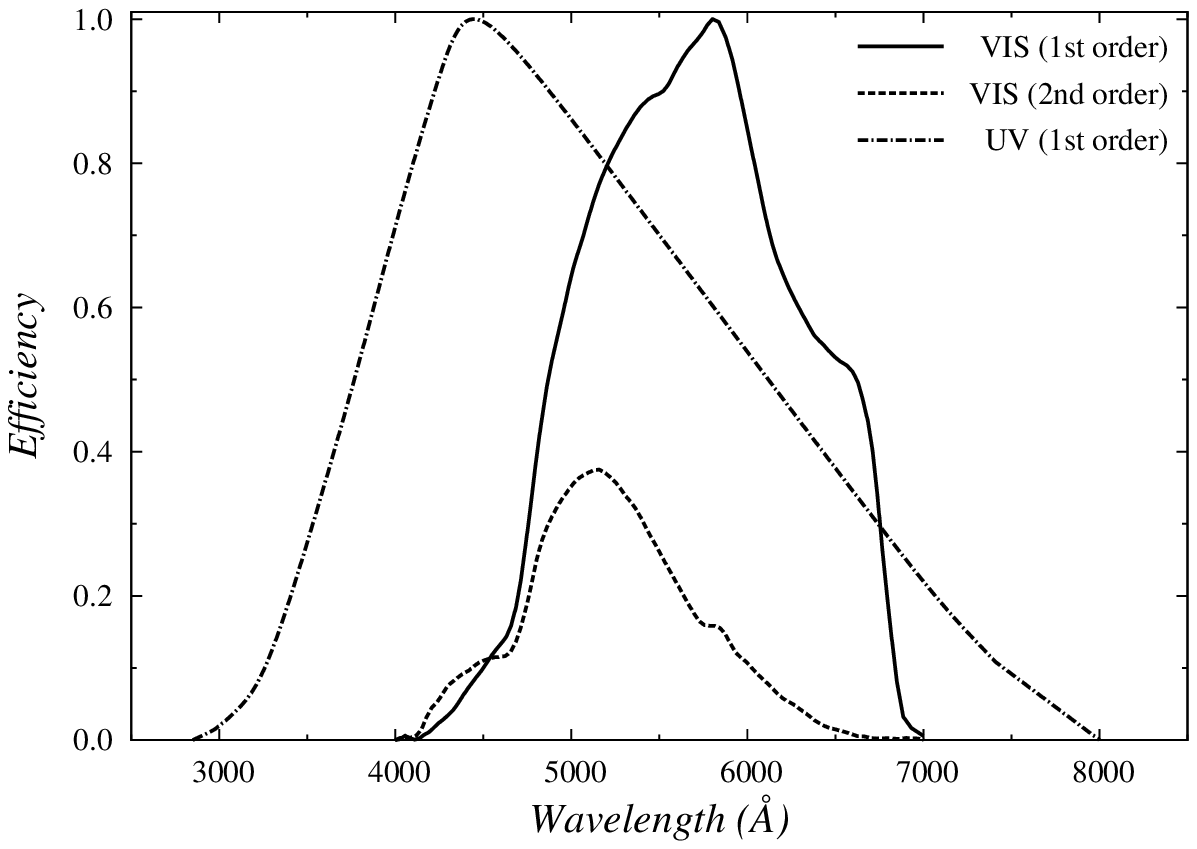}
  \end{center}
  \caption{Efficiency curves of the VIS-HDTV and UV-II spectroscopic cameras. The 1st order spectra are normalized 1.0 at their maximum. The efficiency of the 2nd order of the VIS-HDTV is relative to its 1st order spectrum. The maximum efficiency of the UV-II and the VIS-HDTV's 1st and 2nd orders are 4440, 5804, and 5160 \AA, respectively.}\label{eff}
\end{figure}

A ground observation team consisting of 16 members was organized by JAXA (\cite{fujita11a}). Triangulation observations were coordinated between GOS3 station at Tarcoola (134.55858 E, --30.699114 S, 0.152 km altitude) and GOS4 station at Coober Pedy (134.71819 E, --29.03392 S, 0.224 km altitude) in southern Australia (figure~\ref{map}). The National Astronomical Observatory of Japan (NAOJ) also made a HAYABUSA observing expedition at Coober Pedy (see \cite{watanabe11}).

Spectroscopy was carried out using two spectrograph systems: a VISual HDTV camera (VIS-HDTV) and an UltraViolet Image-Intensified TV camera (UV-II). An EOS 5D MarkII with lens (f=80mm, f/5.6) was applied to the VIS-HDTV system equipped with a transmission grating (300 grooves per mm, blazed at 6100 \AA). The UV-II system consisted of a UV lens (f=30mm, f/1.2), a UV image intensifier ($\phi$ 18-mm photo-cathode: S20), two relay lenses (f=50mm, f/1.4), and a SONY HDTV Handycam. The UV-II system was equipped with a reflection grating (600 grooves per mm, blazed at 3300 \AA). The VIS-HDTV camera obtained the 0th, 1st, and 2nd order spectra in 4000 -- 7000 \AA~, and the UV-II camera observed the 1st order spectrum in 3000 -- 7500 \AA~ (figure~\ref{eff}). A part of the 2nd order spectrum of the VIS-HDTV was overlapped with the 3rd order spectrum. The specifications of the spectroscopic and imaging cameras are summarized in table \ref{tb_instrument}.

The tracking observations at GOS3 were carried out by S. Abe using co-aligned dual imaging cameras (Nikon D300s and WAT-100N) and the UV-II spectrograph which were mounted on the same hydraulic tripod. Tracking was performed while watching a video monitor taken by WAT-100N (WATEC Inc.). The high-sensitivity television camera, NCR-550a (NEC Corp. \& GOTO Inc.), equipped with three 1/2 type EM-CCD (Electron Multiplying CCD) image sensors, was also used to observe the zoomed color TV operated by O. Iiyama. At GOS4, the VIS-HDTV spectroscopy was achieved by Y. Kakinami adopting the same tracking method. Meanwhile, 6$\times$7 photographic cameras using a fish-eye lens with a rotating shutter were operated simultaneously at GOS3 by S. Abe and at GOS4 by Y. Kakinami and Y. Shiba. In this paper, the most reliable trajectory (time, height, and velocity) determined by our photographic observation is referred (\cite{borovicka11}). Note that the trajectories derived from different cameras (\cite{shoemaker11}; \cite{ueda11}) were comparable to our result and the JAXA prediction. Tracking the capsule was difficult because although there were predictions for the time and point in the sky when the objects would first appear, the capsule moved faster and tracking became more difficult later in flight. In order to track the fast moving HAYABUSA trajectory smoothly, observers were sufficiently trained by using an imitation moving laser pointer on the planetarium dome that was arranged by M. Suzuki. These instrument pointing exercises allowed successful tracking of the HAYABUSA emissions (\cite{abe10}).

Background and stars were removed by subtracting a median frame shortly before or after the spectrum. After flat-fielding and averaging of the HAYABUSA spectrum, the wavelength was examined carefully using well-known strong atomic lines such as Mg\emissiontype{I} (5178 \AA) and Na\emissiontype{I} (5893 \AA). The pixels were converted to wavelengths with a simple linear function. After more known lines were identified, the wavelength was precisely determined again with regard to the synthetic atomic lines that were convolved by the spectrum resolution (\cite{borovicka93}). On the other, spectra of Venus and Canopus ($\alpha$ Car) were used to calculate the efficiency curves for the VIS-HDTV and UV-II cameras, respectively (figure~\ref{eff}). Table {\ref{tb_event}} gives trajectories of the HAYABUSA spacecraft (H $>$50 km) and the capsule (H$<$50 km) compared with detections by our imaging and spectroscopic cameras. The VIS-HDTV spectroscopy was aimed at the HAYABUSA spacecraft (50$<$H$<$85 km), whereas the UV-II spectroscopy was intended to observe the faint spectrum at the beginning height (H$>$80 km) and the capsule spectrum near the terminal height (H$<$50 km). Here, we selected some of the best data among a series of spectrum (see figure 12 in \cite{fujita11a}).

\section{Results}

\subsection{VIS-HDTV spectrum in 4000 -- 7000 \AA}

\begin{figure*}
  \begin{center}
    \FigureFile(\linewidth,1mm){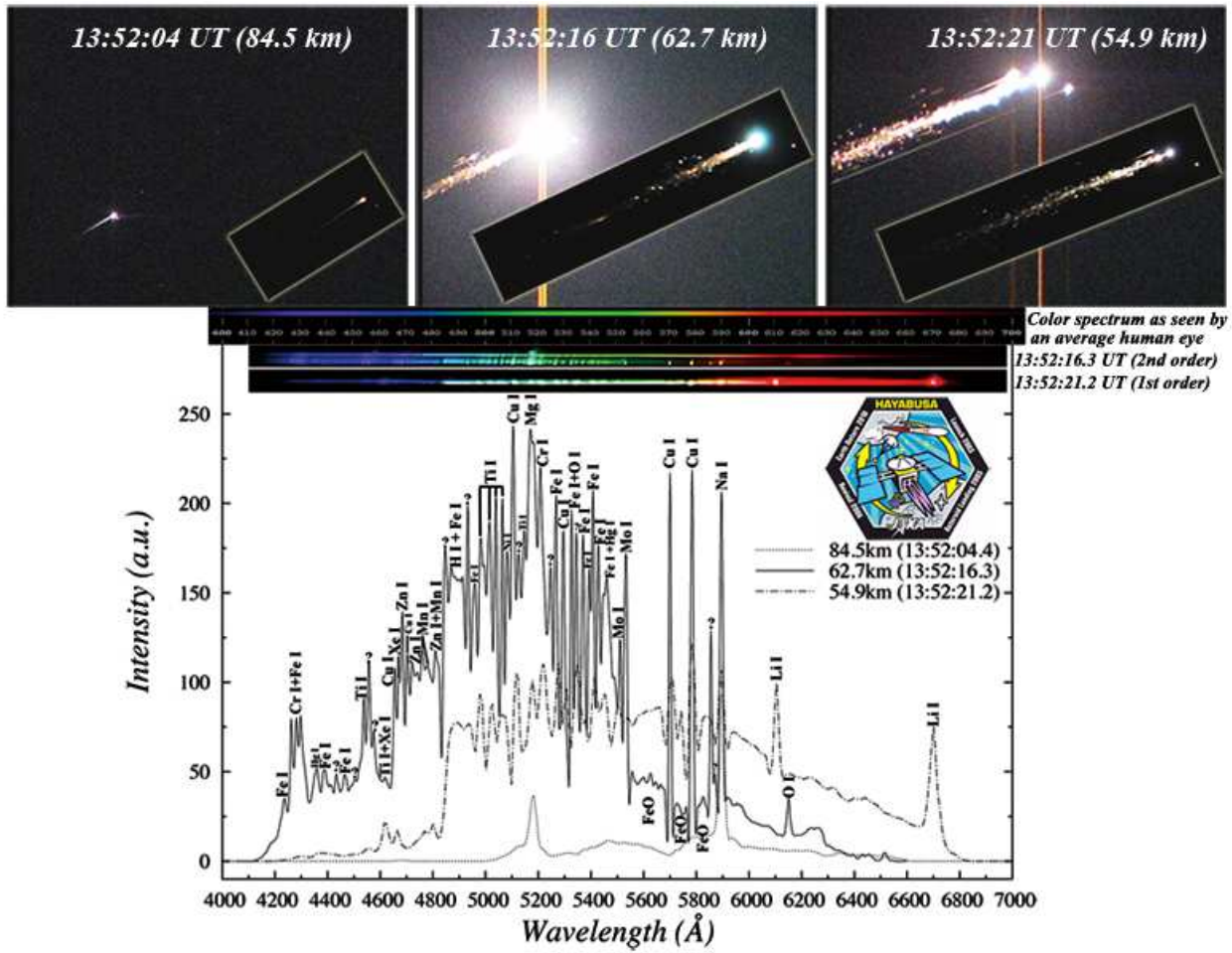}
  \end{center}
  \caption{Visible spectrum of HAYABUSA spacecraft compared with color images taken by the NCR-550a (NEC Corp. \& GOTO Inc.) forming a background and the Nikon D300s in a rectangular box. Identified atomic lines are indicated atop each emission. '?' marks are unknown (unidentified) lines. A color spectrum as seen by an average human eye is shown (Nick Spiker; http://www.repairfaq.org/sam/repspec/).}\label{hayabusa_color}
\end{figure*}

\begin{figure}
  \begin{center}
    \FigureFile(\linewidth,1mm){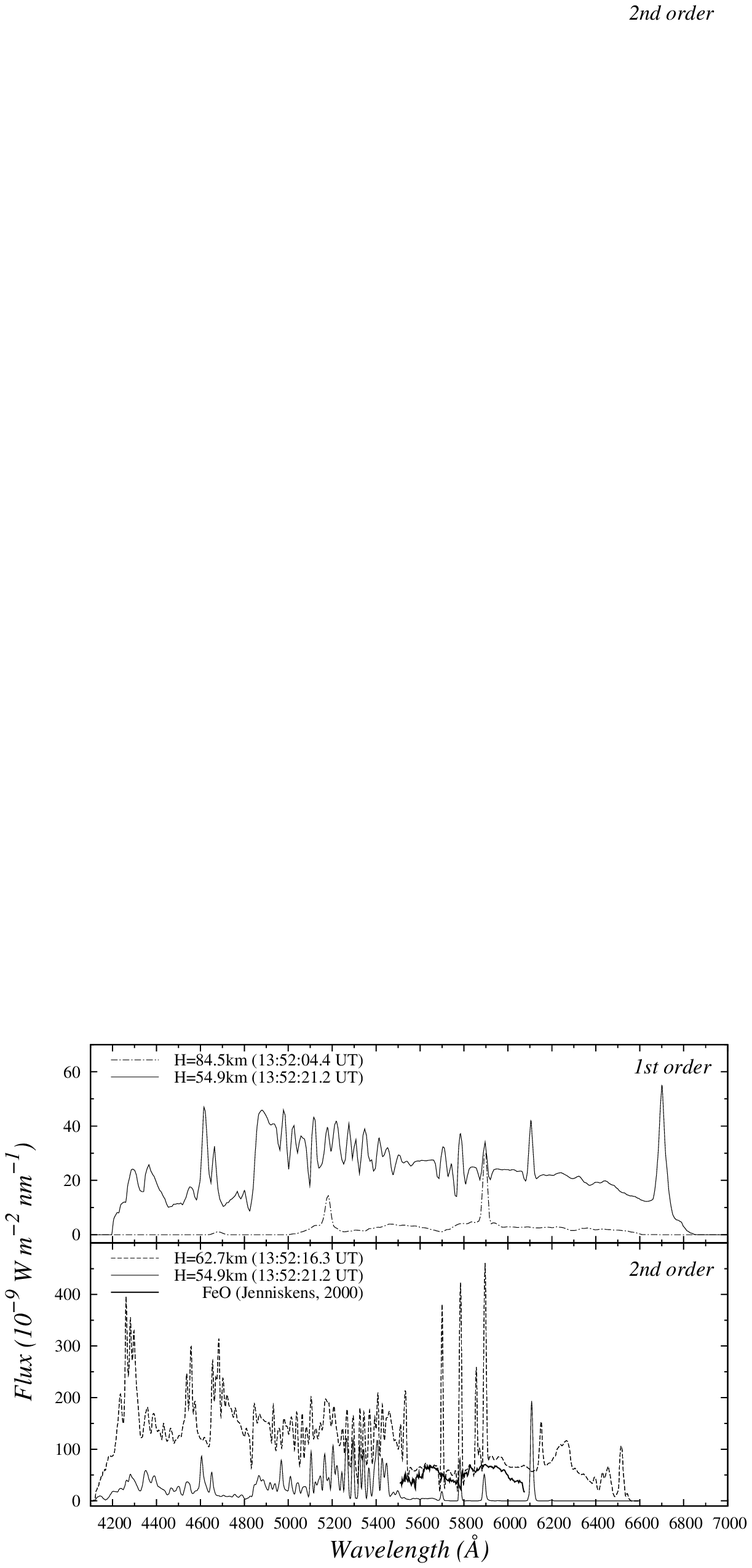}
  \end{center}
  \caption{The VIS-HDTV spectra of the HAYABUSA spacecraft in the 1st (upper panel) and in the 2nd (lower panel) order after spectral sensitivity calibration. The 2nd order spectrum was not obtained for H=84.5 km due to its faintness, while the 1st order spectrum was mostly saturated for H=62.7 km. The spectral match with the laboratory spectrum of the FeO orange bands (\cite{jenniskens00}) is superposed. Since the 3rd order is clearly mixed with the 2nd order spectrum below $\sim$6600 \AA~, the 3rd order spectrum is omitted.}\label{vis_calib}
\end{figure}

Figure~\ref{hayabusa_color} shows the VIS-HDTV spectra of the HAYABUSA spacecraft at heights of 84.5, 62.7, and 54.9 km. The estimated absolute magnitude at each height were --7, --12, and --8, respectively. Note that the absolute magnitude is defined as the magnitude an object would have as if placed at the observer's zenith at a height of 100 km. Corresponding color images taken with the NCR-550a and the Nikon D300s were compared. These spectra were not calibrated by the sensitivity curve so that they could be compared with the color as seen by an average human eye as well as with the color images. Identified atoms were indicated on the top of each emission. After the spectral response calibration, the 1st and the 2nd order spectra were obtained (figure~\ref{vis_calib}). Most of the 1st order spectra of the VIS-HDTV and the UV-II during the explosion phase were saturated in their intensities. The 2nd order spectra of the VIS-HDTV, fortunately, were free from saturation and were worthy of close inspection. The absolute flux was estimated using the unsaturated 2nd order spectrum when the maximum magnitude of --12.6 was reached at a height of 66.9 km (\cite{borovicka11}). Identified atoms and molecules using the VIS-HDTV are summarized in table \ref{tb_identifyVIS}.

At a height of 84.5 km, Mg\emissiontype{I} (5173 and 5184 \AA) and Na\emissiontype{I} (5890 and 5896 \AA) emissions were dominant. During the explosion phase, numerous strong emissions were seen in the visible spectrum. Some exotic lines were detected, such as Cu\emissiontype{I} (5700 and 5782 \AA), Mo\emissiontype{I} (5506 and 5533 \AA), Xe\emissiontype{I} (4624 and 4671 \AA), and Hg\emissiontype{I} (4358 and 5461 \AA), which typically could not be seen in a natural meteor spectrum. Note that the ``duralumin'' of HAYABUSA's structure contains Al, Cu, Mg, and Mn. MoS$_2$ (molybdenum disulfide) was used as a lubricant in many places of the spacecraft. HAYABUSA's propulsion system operated by accelerating ionized Xe (xenon gas) through a strong electric field, and expelling it at high speed. Of the total 66 kg of xenon gas that was carried on HAYABUSA, there remained about 10 kg at the time of the Earth return. Xe\emissiontype{I} at 4671 \AA~ was clearly seen in the spectrum (H=62.7 km) after the main explosion, and disappeared at a lower height (H=54.9 km). The strong Li\emissiontype{I} lines (6104 and 6708 \AA) at a height of 54.9 km were detected from the spectrum of the HAYABUSA spacecraft. It is most likely that the observed Li\emissiontype{I} emissions originated from the Li-Ion batteries consisting of 11 prismatic cells with $\sim$6.3 kg total mass onboard the HAYABUSA spacecraft. A series of strong Zn\emissiontype{I} lines (4680, 4722, and 4811 \AA) were detected that was probably originated from the spacecraft. Similar Zn\emissiontype{I} lines have been seen in a ``paint'' spectrum of the NASA Stardust capsule due to paint that was applied to the surface of the capsule (\cite{abe07a}; \cite{jenniskens10a}).

The continuum spectral profile around 6000 \AA~ at a height of 62.7 km is very similar to published laboratory measurements of the ``orange bands'' of FeO which have been detected by \citet{jenniskens00} and \citet{abe05a} in Leonid meteor persistent trains. FeO is the most common molecule observed in the spectrum of bright and relatively slow fireballs (\cite{ceplecha71}; \cite{borovicka93}). The FeO can be formed during the cooling phase when the temperature drops to 2500 -- 2000 K (\cite{berezhnoy10}), thus it may be emitted in the wake of the HAYABUSA spacecraft. Recently, FeO has been discovered in a terrestrial night airglow spectrum observed with the Odin spacecraft (\cite{evans10}).

\subsection{UV-II spectrum in 3000 -- 4000 \AA}

\begin{table*}
  \caption{Identification of atoms and molecules in 3000 -- 4000 \AA. Identified atoms here were based on the spacecraft-capsule mixed spectrum at a height of 82.9 km (13:52:05 UT), while molecular bands resulted from the spacecraft-capsule mixed spectra at a height of 92.5 km (13:52:00 UT), 82.9 km (13:52:05 UT), and 62.7 km (13:52:16 UT). Identified atoms and synthetic molecular spectra are shown in the UV-II spectrum (figure~\ref{uv_cal}). The precision of the temperature estimation is about $\pm$500 K.}\label{tb_identifyUV}
  \begin{center}
    \begin{tabular}{rcccccc}
\hline
\multicolumn{7}{c}{Identified line} \\
 Band head        & \multicolumn{2}{c}{Temperature}    & \multicolumn{4}{c}{Molecule} \\
 Wavelength (\AA) & \multicolumn{2}{c}{$T_v=T_r$ ($K$)} &                              \\ \hline
 3092 & \multicolumn{2}{c}{2000}  & \multicolumn{4}{c}{OH system ($A^{2}\Sigma^{+}\rightarrow X^{2}\Pi$)}                               \\
 3294 & \multicolumn{2}{c}{4000}  & \multicolumn{4}{c}{N$_{2}^{+}$ ($1^-$) system ($B^{2}\Sigma_{u}^{+} \rightarrow X^{2}\Sigma_{g}^{+}$)}  \\
 3560 & \multicolumn{2}{c}{4000}  & \multicolumn{4}{c}{N$_{2}^{+}$ ($1^-$) system ($B^{2}\Sigma_{u}^{+} \rightarrow X^{2}\Sigma_{g}^{+}$)}  \\
 3908 & \multicolumn{2}{c}{4000}  & \multicolumn{4}{c}{N$_{2}^{+}$ ($1^-$) system ($B^{2}\Sigma_{u}^{+} \rightarrow X^{2}\Sigma_{g}^{+}$)}  \\
 3292 & \multicolumn{2}{c}{13000} & \multicolumn{4}{c}{N$_{2}^{+}$ ($1^-$) system ($B^{2}\Sigma_{u}^{+} \rightarrow X^{2}\Sigma_{g}^{+}$)}  \\
 3533 & \multicolumn{2}{c}{13000} & \multicolumn{4}{c}{N$_{2}^{+}$ ($1^-$) system ($B^{2}\Sigma_{u}^{+} \rightarrow X^{2}\Sigma_{g}^{+}$)}  \\
 3880 & \multicolumn{2}{c}{13000} & \multicolumn{4}{c}{N$_{2}^{+}$ ($1^-$) system ($B^{2}\Sigma_{u}^{+} \rightarrow X^{2}\Sigma_{g}^{+}$)}  \\
 3585 & \multicolumn{2}{c}{13000} & \multicolumn{4}{c}{CN violet system ($B^{2}\Sigma^{+} \rightarrow X^{2}\Sigma^{+}$)}                 \\
 3849 & \multicolumn{2}{c}{13000} & \multicolumn{4}{c}{CN violet system ($B^{2}\Sigma^{+} \rightarrow X^{2}\Sigma^{+}$)}                 \\
\hline
\multicolumn{7}{c}{Identified line} \\
Wavelength (\AA) & Element &  multiplet & & Wavelength (\AA) & Element &  multiplet \\
\cline{1-3}
\cline{5-7}
 3020 & Fe\emissiontype{I} &    9  &  & 3434 & Ni\emissiontype{I} &   19  \\
 3021 & Fe\emissiontype{I} &    9  &  & 3441 & Fe\emissiontype{I} &    6  \\
 3091 & Mg\emissiontype{I} &    5  &  & 3441 & Fe\emissiontype{I} &    6  \\
 3092 & Fe\emissiontype{I} &   28  &  & 3444 & Fe\emissiontype{I} &    6  \\
 3093 & Al\emissiontype{I} &    3  &  & 3446 & Ni\emissiontype{I} &   20  \\
 3093 & Al\emissiontype{I} &    3  &  & 3648 & Fe\emissiontype{I} &   23  \\
 3134 & Ni\emissiontype{I} &   25  &  & 3687 & Fe\emissiontype{I} &   21  \\
 3226 & Fe\emissiontype{I} &  155  &  & 3944 & Al\emissiontype{I} &    1  \\
 3233 & Ni\emissiontype{I} &    7  &  & 3962 & Al\emissiontype{I} &    1  \\
 3415 & Ni\emissiontype{I} &   19  &  & 3969 & Fe\emissiontype{I} &   43  \\
 3424 & Ni\emissiontype{I} &   20  &  &      &                    &       \\
\hline
    \end{tabular}
  \end{center}
\end{table*}

\begin{figure}
  \begin{center}
    \FigureFile(\linewidth,1mm){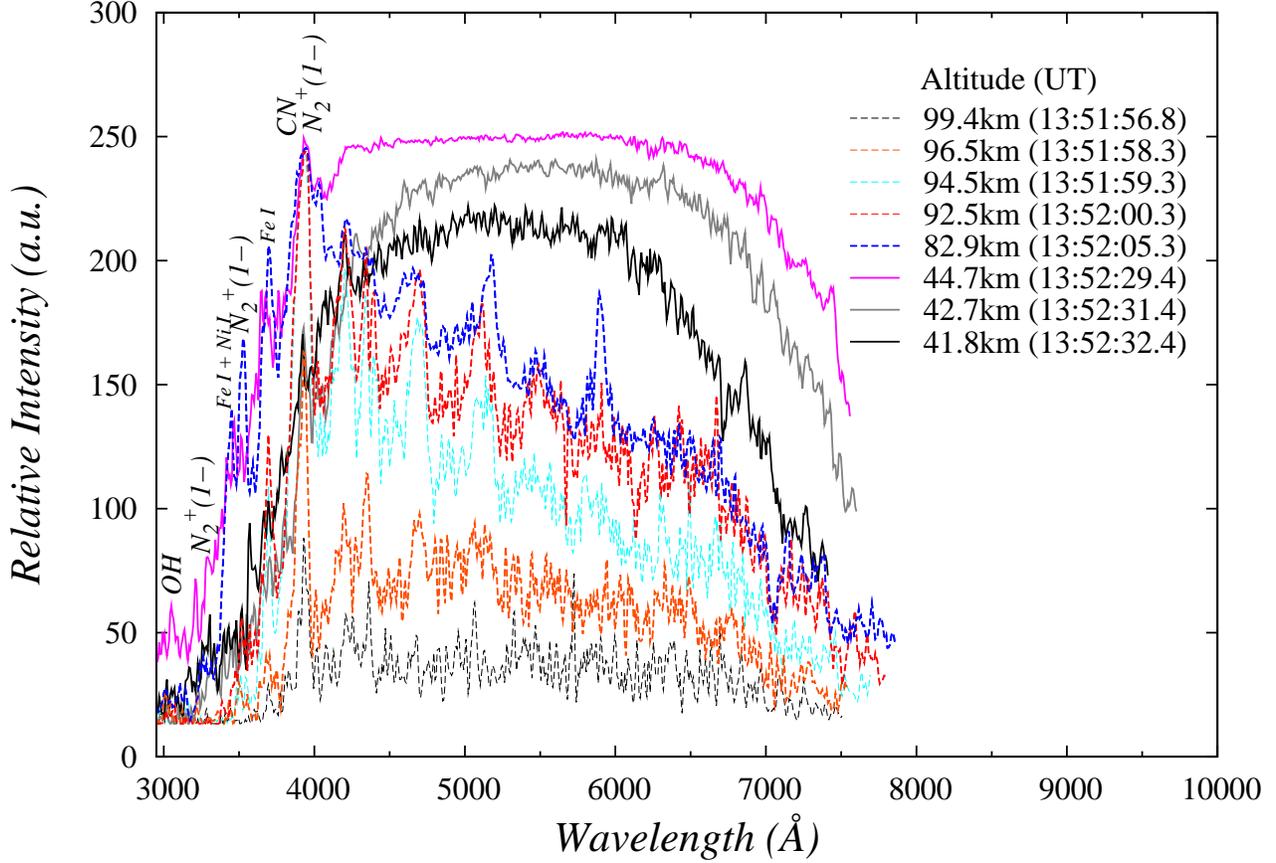}
  \end{center}
  \caption{A time series of the UV-II spectra of HAYABUSA spacecraft (dotted lines) and the re-entry capsule (solid lines) before spectral sensitivity calibration. The saturation in intensity reaches near 250 in this figure. Some important molecular and atomic species below 4000 \AA~ are indicated atop each emissions.}\label{uv_raw}
\end{figure}

\begin{figure}
  \begin{center}
    \FigureFile(\linewidth,1mm){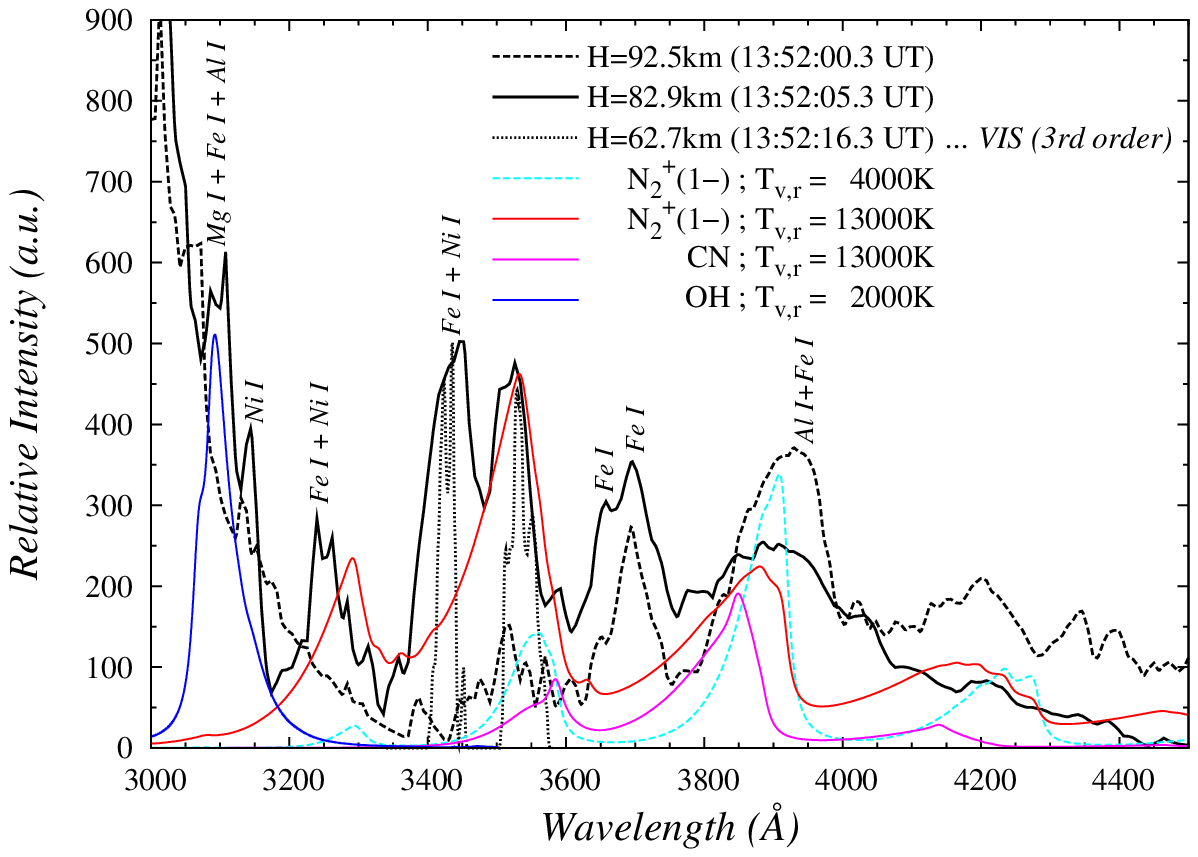}
  \end{center}
  \caption{Selected UV-II spectra of HAYABUSA spacecraft after sensitivity calibration compared. The scaled 3rd order of the VIS-HDTV spectrum without sensitivity calibration is shown in the 3400 -- 3600 \AA~ range. N$_{2}^{+}$ ($1^-$) bands from a shock layer and CN violet bands from an ablating heat shield of a sample return capsule are dominant molecular bands. OH($A$-$X$) band is likely to exist around 3092 \AA, but blended with atomic lines. Assuming that a rotation temperature equals to a vibration temperature, the model spectrum of N$_{2}^{+}$ ($1^-$), CN, and OH($A$-$X$) systems are superposed.}\label{uv_cal}
\end{figure}

Figure {\ref{uv_raw}} shows a series of the UV-II spectra for the spacecraft-capsule mixed at heights between 99.4 km -- 82.9 km and for the capsule at heights between 44.7 km -- 39.6 km before spectral sensitivity calibration. N$_{2}^{+}$ ($1^-$) at 3908 \AA~ was a significant band head during the beginning phase, and during the later phase the near-ultraviolet region was filled with N$_{2}^{+}$ ($1^-$) bands whose band heads were 3880 and 3533 \AA. Figure \ref{uv_cal} shows the UV-II spectra at heights of 92.5 km  and 82.9 km after sensitivity calibration. A scaled 3rd order spectrum obtained using the VIS-HDTV is superposed on this figure. The model spectrum of N$_{2}^{+}$ ($1^-$) ($B^{2}\Sigma_{u}^{+} \rightarrow X^{2}\Sigma_{g}^{+}$) system with four bands heads (3300, 3500, 3900 and 4200 \AA) caused by different vibrational states were carried out varying the temperatures from 1000 K to 20000 K using the SPRADIAN numerical code (\cite{fujita97}). Assuming that a rotation temperature equals to a vibration temperature, N$_{2}^{+}$ ($1^-$) bands convolved by the spectral resolution (20 \AA) were examined (figure~\ref{uv_cal}). The CN violet bands ($B^{2}\Sigma^{+}\rightarrow X^{2}\Sigma^{+}$) and a possible contribution of OH band ($A^{2}\Sigma^{+}\rightarrow X^{2}\Pi$) were also demonstrated in the same way. N$_{2}^{+}$ molecules were originated from the atmospheric gas. CN molecules were produced by a chemical product of ablated carbon atoms from the heat shield of the capsule and atmospheric nitrogen molecules. Both N$_{2}^{+}$ and CN molecules were generated in the shock layer of the body. The derived temperatures of N$_{2}^{+}$ ($1^-$) bands at heights of 92.5 km and 82.9 km were $\sim$4000 K and $\sim$13000 K, respectively. Though most of CN bands were buried under strong N$_{2}^{+}$ ($1^-$) bands, a clear CN band head at 3533 \AA~ was detected which is explained by the vibration temperature of $\sim$13000 K. The other emission features consisted mainly of Fe\emissiontype{I}, Mg\emissiontype{I}, Al\emissiontype{I} and Ni\emissiontype{I}. Table~\ref{tb_identifyUV} gives the identification of atoms and molecules in the range between 3000 -- 4000 \AA.

\subsection{Gray-body spectrum of the capsule}

\begin{table}
  \caption{Gray-body temperature of the sample return capsule.}\label{tb_gray}
  \begin{center}
    \begin{tabular}{cccc}
\hline
    Time    & Height & Velocity  & Temperature   \\
     UT     &  $km$  & $km s^{-1}$ &  $K$         \\ \hline
13:52:31.35 & 42.69  & 6.35      & 2482 $\pm$ 11 \\
13:52:32.35 & 41.81  & 5.88      & 2437 $\pm$ 14 \\
13:52:32.99 & 41.28  & 5.58      & 2395 $\pm$ 15 \\
\hline
    \end{tabular}
  \end{center}
\end{table}

\begin{figure}
  \begin{center}
    \FigureFile(\linewidth,1mm){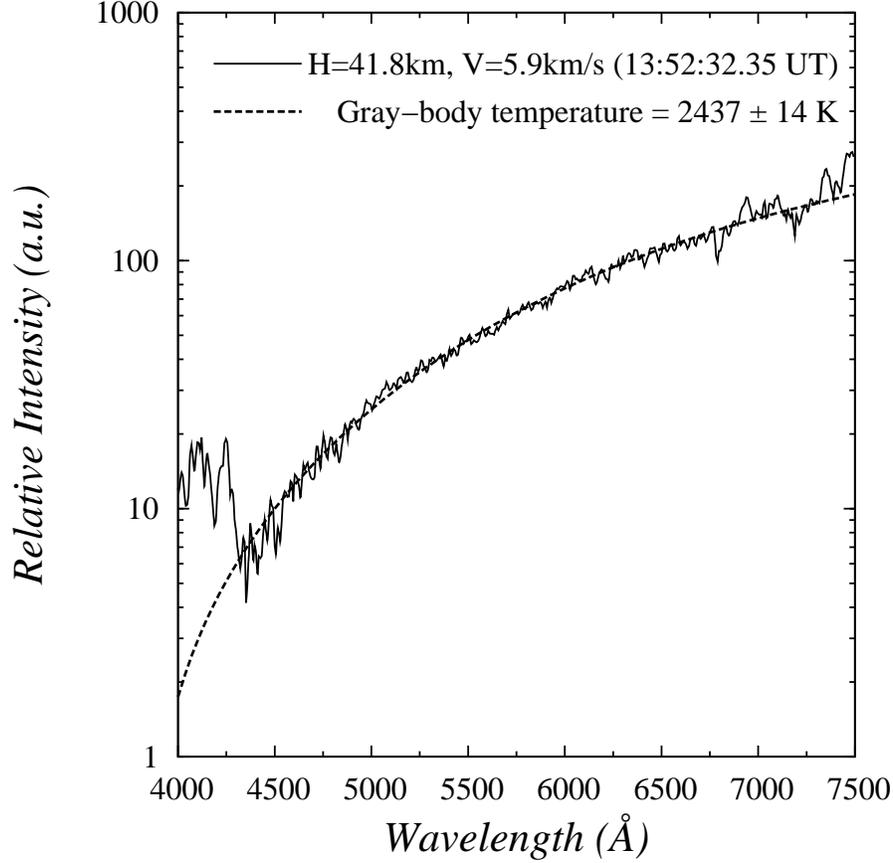}
  \end{center}
  \caption{The gray body spectra of the sample return capsule at a height of $\sim$42 km. Near-ultraviolet region below 4500 \AA~ is affected by atom and molecule emissions. Thus, Plank's formula was applied to the range between 4500 -- 7000 \AA~ as shown.}\label{gray}
\end{figure}

A black-body is an idealized object with a perfect absorber and emitter of radiation whose emissivity is larger than 0.99. An object which has lower emissivity such as a re-entry capsule is often referred to as a gray-body. The capsule spectrum was well separated from HAYABUSA's fragments below $\sim$45 km in height while N$_{2}^{+}$ ($1^-$) and CN bands were still strong in the near-ultraviolet region (figure~\ref{uv_raw}). After sensitivity calibration, the gray-body temperature in the wavelengths between 4500 -- 6500 (7000) \AA~was estimated employing Planck's formula. The gray-body temperatures of 2482 $\pm$ 11 K, 2437 $\pm$ 14 K, and 2395 $\pm$ 15 K were measured at heights of 42.7 km,  41.8 km, and 41.2 km, respectively (table~\ref{tb_gray}; figure~\ref{tb_gray}).

\section{Discussion}

\subsection{VIS-HDTV spectrum in 4000 -- 7000 \AA}

\begin{table*}
  \caption{Derived excitation temperature of radiating gas, the column density and the intensities of identified main neutral atoms. The precision of the temperature estimation is about $\pm$500 K, and that of other parameters is within a factor of 2--3.}\label{tb_element}
  \begin{center}
    \begin{tabular}{ccccc}
  \hline
Height ($km$)  &   84.51    &   62.72    &  54.94     &   54.94   \\
Spectrum order &    1st     &    2nd     &    1st     &    2nd    \\
T ($K$)        &   6000     &   6000     &   4500     &    4500   \\ \hline
%$N (cm^{-2})$\footnotemark[$\dagger$]
%               &   2.0E+13  &   1.0E+13  &  1.0E+13   &   1.0E+13 \\
Mg\emissiontype{I}           &   1.0E+0   &   1.0E+0   &  1.0E+0    &   1.0E+0  \\
Fe\emissiontype{I}           &   6.0E-1   &   1.0E+1   &  9.0E+0    &   9.0E+0  \\
Mn\emissiontype{I}           &    --      &   2.0E-1   &  3.0E-1    &   3.0E-1  \\
Ti\emissiontype{I}           & $<$1.0E-3  &   2.0E-2   &  2.0E-2    &   2.0E-2  \\
Na\emissiontype{I}           &   1.2E-2   &   1.8E-2   &  1.0E-3    &   7.0E-4  \\
Cr\emissiontype{I}           &   1.2E-2   &   4.0E-2   &  4.0E-2    &   4.0E-2  \\
Li\emissiontype{I}           &     --     &     --     &  3.2E-3    &     --    \\
Ni\emissiontype{I}           & $<$1.0E+1  &   1.0E+1   & $<$1.0E+0  &   5.0E+0  \\
%O\emissiontype{I}            & $<$6.0E+6  &   3.0E+7   & $<$1.0E+7  & $<$1.0E+7 \\
%N\emissiontype{I}            & $<$2.0E+7  &   5.0E+7   & $<$1.0E+9  & $<$1.0E+9 \\
  \hline
    \end{tabular}
  \end{center}
%\footnotemark[$\dagger$] The column density is measured in units of number per area.
\end{table*}

When the surface temperature of HAYABUSA reached about 2000 K, occurring at a height around 100 km, the spacecraft material started to sublimate from the surface and was surrounded by evaporated vapors. Excited states of atoms of these vapors were gradually de-excited by radiation. HAYABUSA luminosity consisted mostly of radiation of discrete emission spectral lines belonging under the most part to metals and mainly to iron. Ablation particles injected into the flow also emitted radiation as a continuum spectrum due to their temperature. Thus, observed atomic spectrum was mixed with thermal continuum and molecular bands. Subtracting these background, excitation temperature and element intensities were estimated under the assumption of a local thermal equilibrium (LTE) condition. Table~\ref{tb_element} gives the derived physical quantities. 

The excitation temperature was estimated using Fe\emissiontype{I} and Mg\emissiontype{I} lines around the 5000 -- 5500 \AA wavelength range. The derived excitation temperatures were 6000 K at heights of 84.5 km and 62.7 km, and 4500 K at a height of 54.9 km. These excitation temperatures are comparable to a normal meteor temperature, $\sim$5000 K (e.g, \cite{borovicka93}; \cite{ceplecha96}; \cite{trigo03}) and the excitation temperature range between 5000 K -- 6000 K caused by a large bolide (\cite{borovicka98}). Although the ratio of Fe\emissiontype{I}/Mg\emissiontype{I} was 0.6 at a height of 84.5 km, the ratio increased to 10 below a height of 62.7 km. Since HAYABUSA was made with much more Fe than Mg, it is likely that iron was conspicuously evaporated during the explosion phase.

Detected Ti\emissiontype{I} lines seem to originate from fuel and gas tanks, and also bolts with nuts that were made of Ti-6Al-4V. H\emissiontype{I} line (4861 \AA) is possibly present, presumably as a result of the dissociation of N$_2$H$_4$ which was used as thruster fuel, meaning that some of the thrusting fuel remained at re-entry. All other elements (Ni, Cr, and Mn) were probably used for constructing the HAYABUSA spacecraft. Though Hg\emissiontype{I} was identified as the most probable element to explain the presence of lines at 4358 and 5461 \AA, details are proprietary information that were not available to the authors. The detected Na\emissiontype{I}, especially at heights of 62.7 km and 54.9 km, is not likely to have originated from the atmosphere; although Earth's atmosphere contains natural sodium known as the sodium layer at a height of 80 -- 105 km which originated from meteoroids, this sodium is more rare at these low altitudes. Therefore, sodium must have originated from material ablated by the HAYABUSA spacecraft or its capsule. Sodium was also detected by the Stardust capsule at altitudes of 54 -- 48 km (\cite{nielsen10}). 

O\emissiontype{I} and N\emissiontype{I} are most likely atmospheric lines, or a part of O\emissiontype{I} and N\emissiontype{I} arising from the fuel oxidizer, N$_2$O$_4$. O\emissiontype{I}, N\emissiontype{I}, H\emissiontype{I}, and Xe\emissiontype{I} are of high excitation atoms which require either high temperature or a large amount of atoms to be detected. Therefore, these atoms are thought to be excited by the shock layer. On the other hand, typical lines of the high temperature components as seen in the meteor plasma such as Mg\emissiontype{II} (4481 \AA) and Fe\emissiontype{II} (e.g., 4583, 4923, and 5018 \AA) have not been detected in HAYABUSA spectra. The one explanation for the absence of Mg\emissiontype{II} and Fe\emissiontype{II} is that the emissions of O\emissiontype{I}, N\emissiontype{I}, H\emissiontype{I}, and Xe\emissiontype{I} arise promptly from the shock layer in the gas--gas phase, while the ablation of solid Mg and Fe should transform into a gas phase that takes more time than gas--gas transformation because of the low velocity ($\sim$10 km/s) compared with typical meteors ($\sim$40 km/s).

\subsection{UV-II spectrum in 3000 -- 4000 \AA}

The vibration temperature of molecular N$_{2}^{+}$ ($1^-$) was dramatically changed from 4000 K at a height of 92.5 km to 13000 K at a height of 82.9 km (table~\ref{tb_identifyUV}; figure~\ref{uv_cal}). The observed spectra are a superposition of the post shock plasma radiation which is mixed with a shock layer heating and downward plasma. Thus, it is logical to understand that the high temperature region was induced by a shock layer of the HAYABUSA spacecraft which rapidly grew between 92.5 km and 82.9 km in height. The molecular band of N$_{2}^{+}$ ($1^-$) was also observed in the spectrum of the Stardust re-entry capsule with a rotational temperature of 15000 K at heights between 71.5 km and 62 km (\cite{winter11}). Interestingly, at a height of 84 km, N$_{2}^{+}$ ($1^-$) was detected with a vibration temperature of $\sim$10000 K from a --4 magnitude bright fireball of the Leonid meteor shower whose entry velocity was $\sim$72 km/s (\cite{abe05b}). The flux of the spacecraft at a height of 62.7 km (--12 absolute magnitude) was about 1600 times brighter than that of the capsule's flux (--4 absolute magnitude at the maximum). Therefore, the N$_{2}^{+}$ ($1^-$) bands originated from the spacecraft was much stronger than CN bands originated from the capsule in which C was the major erosion product of the Carbon-Phenol heat shield of the capsule as seen by the Stardust capsule (\cite{jenniskens10a}; \cite{winter11}). Clear C\emissiontype{I} lines were also observed in the near-infrared spectrum (around 1 $\mu$ m wavelength) of the Stardust reentry capsule (\cite{taylor10}). Note that the vibrational and rotational temperatures of CN violet bands were measured to be 8000 $\pm$ 1000 K in the Stardust capsule at a height of 60 km (\cite{jenniskens10b}). C--N coupling occurs at a higher excited state than a ground state, and then vibration-rotation temperature approaches to the translational temperature. For instance, the estimated translation temperature in the shock layer of the capsule was 11000--13000 K. Hence, the vibration-rotation temperature of $\sim$13000 K for N$_{2}^{+}$ ($1^-$) and CN is reasonable.

Bright fireballs sometimes leave a self-luminous long-lasting plasma at altitudes of about 80--90 km that is called 'persistent trains'. It is generally believed that the luminosity of persistent trains is fueled by reactions involving O$_{3}$ and atomic O, efficiently catalyzed by metals from the freshly ablated meteoroids (e.g., \cite{jenniskens00}; \cite{abe05a}). A persistent train was observed at heights between 92 km and to 82 km for about 3 minutes(\cite{yamamoto11}). It is reasonable to suppose that sufficient metals were supplied by the ablation of the spacecraft at these heights.

It is important to understand how meteoroids and meteors supply the Earth with space matter including organics and water (e.g., \cite{abe07b}). Meteors represent a unique chemical pathway towards prebiotic compounds on the early Earth and a significant fraction of organic matter is expected to survive. Thus, the investigations of OH($A$-$X$) in the HAYABUSA spectrum is of particular interest. The most likely mechanism for emitting OH($A$-$X$) band in the meteor is caused by the dissociation of water or mineral water in the meteoroid. Our possible detection of OH($A$-$X$) band indicates an Earthly origin caused by the dissociation of water in the upper atmosphere. However, due to blending with other atomic lines such as Mg\emissiontype{I}, Fe\emissiontype{I}, and Al\emissiontype{I}, further spectroscopy with higher resolution and sensitivity around 3090 \AA~ will be needed for further confirmation.

\subsection{Gray-body spectrum of the capsule}

A continuum consisting of a gray-body emission of the capsule at near-ultraviolet and visible wavelengths was examined. A continuum radiation of the capsule obtained at a height of $\sim$42 km was that of a gray-body at a fitted temperature employing Planck's formula. Our derived temperature from an excellent UV-II data set was 2437 $\pm$ 14 K at a height of 41.8 km (figure~\ref{gray}). The surface temperature of 2525 K $\pm$ 50 K at a height of 41.1 km was estimated based on radiative equilibrium Computational Fluid Dynamics (CFD) (\cite{fujita03}). Our observed temperature agrees qualitatively with the CFD model prediction. The gray-body temperature observed from NAOJ's team at Coober Pedy resulted in $\sim$2400 $\pm$ 300 K at a height of $\sim$40.5 km (\cite{ohnishi11}) which is comparable with our result. Note that dynamical pressure at a height around 40 km was the maximum $\sim$64 kPa (\cite{yamada06}). At the GOS3 site, a strong sequence of sonic booms was detected at 13:55:21 UT using a video recorder's microphone which was observed at Tarcoola by Y. Akita and his colleagues. A sonic boom is the sound associated with a shock wave created by the supersonic flight of the capsule. According to this time delay, the shock wave generation point was estimated at a height around 40 km. Considering these fluid conditions, further analysis will be made in a forthcoming paper (e.g., \cite{fujita11b}).

\section{Conclusion}

At 13:51:50 UT on June 13, 2010, the HAYABUSA spacecraft appeared as planned in the dark sky over the Australian desert, along with the faint dot of the capsule. The HAYABUSA spacecraft was flying behind the capsule, roughly 1 km above as if he must protect his tiger cub. While the spacecraft burst into many fragments, as if falling into the Milky Way, the capsule became an independent bright fireball wearing an ablative heat shield as its thermal protection system (TPS), as if demonstrating its will to overcome adversity. The planned atmospheric re-entry was perfectly completed. The HAYABUSA spacecraft ended his journey in a brilliant flash of light that provided us a treasure trove of ``artificial fireball'' data, which has never been observed in a scientific way. Atomic lines such as Fe\emissiontype{I}, Mg\emissiontype{I}, Na\emissiontype{I}, Al\emissiontype{I}, Cr\emissiontype{I}, Mn\emissiontype{I}, Ni\emissiontype{I}, Ti\emissiontype{I}, Li\emissiontype{I}, Zn\emissiontype{I}, O\emissiontype{I}, and N\emissiontype{I} were identified. The excitation temperature ranging from 4500 K to 6000 K was estimated using Fe\emissiontype{I} and Mg\emissiontype{I} within the 5000 -- 5500 \AA~wavelength region, which is similar to a common excitation temperature of meteors and bolides. The identification of emission lines may be inadequate and contain some unknown lines. Exotic atoms such as Cu\emissiontype{I}, Mo\emissiontype{I}, Xe\emissiontype{I}, and Hg\emissiontype{I} were also identified. The explosion of the spacecraft injected a large amount of iron which increased the density of Fe\emissiontype{I}. The surprising strong red line during the last flash of the spacecraft was well explained by a Li\emissiontype{I} line (6708 \AA) that was probably caused by explosions of the onboard Li-Ion batteries. A clearly detected FeO band at a height of $\sim$63 km is similar to a common bolide spectrum which is likely to emit in the wake, where the radiation is emitted just behind the body. The alteration in the hot plasma temperature of N$_{2}^{+}$ ($1^-$) bands (from 4000 K to 13000 K) appears to be the strongest proof that an intense shock layer around the HAYABUSA spacecraft was rapidly formed at heights between 93 km and 83 km. The gray-body temperatures ranging from 2482 K to 2395 K were measured at heights between 42.7 km and 41.2 km that can be explained by the CFD model prediction. Further investigation is required to understand the performance of the TPS. Our experiences will be instructive in observing the planned HAYABUSA-II Earth return mission.

\bigskip

{\it Acknowledgements.}
We are thankful to all of the ISAS/JAXA personnel who organized and participated in the recovery operations, especially Jun-ichiro Kawaguchi, Hitoshi Kuninaka, Tetsuya Yamada, Makoto Yoshikawa, Masanao Abe, and Hajime Yano. We are also grateful to all of the JAXA ground observation team members at JAXA, NAOJ, Kochi University of Technology, Kanazawa University, and Nagoya University who assisted in observing at GOS3 and GOS4 stations. GOTO Inc. and NEC Corp. kindly provided a newly developed high-sensitive EMCCD camera (NCR-550a). Noboru Ebizuka (Nagoya University) designed the UV-II spectroscopic camera with the aid of SHOWA Industry Co., Ltd in Japan. Yuichiro Akita kindly provided recording data of the sonic boom observed at Tarcoola. We express many thanks to Ji\v{r}\'{i} Borovi\v{c}ka (Ond\v{r}ejov observatory) for constructive comments and kind advice. Synthetic spectrum was carried out in part on the general-purpose PC farm at the Center for Computational Astrophysics (CfCA) of NAOJ. Shinsuke Abe is supported by JAXA and the National Science Council of Taiwan (NSC 97-2112-M-008-014-MY3, NSC 100-2112-M-008-014-MY2). {\it Welcome Back Home ``Okaerinasai'' HAYABUSA!}

%%%%%%%%%%%%%%%%%%%%%%%%%%%%%%%%%%%%%%%

\begin{longtable}{crcccc}
  \caption{Identification of atoms and molecules in 4200 -- 6800 \AA. All atoms except Li\emissiontype{I} in the table were analyzed from the spacecraft spectrum at a height of 62.7 km (13:52:16 UT). Li\emissiontype{I} lines resulted from the spacecraft spectrum at a height of 54.9 km (13:52:21 UT). Identified atoms are indicated in the visual spectrum (figure~\ref{hayabusa_color}). The values for intensities were measured in relative units.}\label{tb_identifyVIS}
  \hline
  \multicolumn{2}{c}{Observed line} & & \multicolumn{3}{c}{Identified line}     \\
  Wavelength (\AA)  & Intensity     & & Wavelength (\AA) & Element &  multiplet \\
  \cline{1-2} \cline{4-6}
\endfirsthead
  \multicolumn{2}{c}{Observed line} & & \multicolumn{3}{c}{Identified line}     \\
  Wavelength (\AA)  & Intensity     & & Wavelength (\AA) & Element &  multiplet \\
  \cline{1-2} \cline{4-6}
\endhead
\endfoot
 4234 &  10.9 & & 4234 & Fe\emissiontype{I} &  152   \\
 4236 &  11.0 & & 4236 & Fe\emissiontype{I} &   52   \\
 4256 &  12.2 & & 4254 & Cr\emissiontype{I} &    1   \\
 4262 &  29.3 & & 4261 & Fe\emissiontype{I} &  152   \\
 4282 &  24.7 & & 4272 & Fe\emissiontype{I} &   42   \\
      &  24.7 & & 4275 & Cr\emissiontype{I} &    1   \\
      &  24.7 & & 4290 & Cr\emissiontype{I} &    1   \\
 4300 &  19.5 & & 4308 & Fe\emissiontype{I} &   42   \\
 4358 &   6.1 & & 4358 & Hg\emissiontype{I} &    1   \\
 4388 &   5.1 & & 4384 & Fe\emissiontype{I} &   41   \\
 4408 &   2.0 & & 4405 & Fe\emissiontype{I} &   41   \\
 4414 &   2.0 & & 4415 & Fe\emissiontype{I} &   41   \\
 4434 &   3.4 & &      &  ?                 &   --   \\
 4466 &   2.8 & & 4467 & Fe\emissiontype{I} &  350   \\
 4502 &   1.6 & &      &  ?                 &   --   \\
 4538 &  13.8 & & 4533 & Ti\emissiontype{I} &   42   \\
 4558 &  19.2 & &      &  ?                 &   --   \\
 4575 &   8.6 & &      &  ?                 &   --   \\
 4618 &   1.8 & & 4617 & Ti\emissiontype{I} &    4   \\
 4622 &   1.6 & & 4624 & Xe\emissiontype{I} &    1   \\
 4656 &  14.6 & & 4651 & Cu\emissiontype{I} &    1   \\
 4670 &  10.2 & & 4671 & Xe\emissiontype{I} &    1   \\
 4684 &  16.0 & & 4680 & Zn\emissiontype{I} &    1   \\
 4704 &   7.3 & & 4705 & Cu\emissiontype{I} &    1   \\
 4724 &   4.0 & & 4722 & Zn\emissiontype{I} &    1   \\
 4540 &   1.7 & &      &  ?                 &   --   \\
 4760 &   2.6 & & 4754 & Mn\emissiontype{I} &   16   \\
 4762 &   2.2 & & 4762 & Mn\emissiontype{I} &   21   \\
 4782 &   1.6 & & 4783 & Mn\emissiontype{I} &   16   \\
 4811 &   3.6 & & 4811 & Zn\emissiontype{I} &    1   \\
 4822 &   4.3 & & 4824 & Mn\emissiontype{I} &   16   \\
 4846 &  12.0 & &      &  ?                 &   --   \\
 4868 &   9.3 & & 4861 & H\emissiontype{I}  &   20   \\
 4872 &   9.0 & & 4872 & Fe\emissiontype{I} &  318   \\
 4898 &   7.0 & & 4891 & Fe\emissiontype{I} &  318   \\
 4912 &   6.4 & &      &  ?                 &   --   \\
 4932 &   9.2 & &      &  ?                 &   --   \\
 4958 &   4.3 & & 4957 & Fe\emissiontype{I} &  318   \\
 4982 &   6.1 & & 4982 & Ti\emissiontype{I} &   38   \\
 4992 &   4.7 & & 4991 & Ti\emissiontype{I} &   38   \\
 5014 &   6.4 & & 5014 & Ti\emissiontype{I} &   38   \\
 5040 &   8.6 & &      &  ?                 &   --   \\
 5081 &   5.8 & & 5081 & Ni\emissiontype{I} &    4   \\
 5016 &   9.4 & & 5106 & Cu\emissiontype{I} &    1   \\
 5125 &   2.7 & &      &  ?                 &   --   \\
 5146 &   3.2 & & 5147 & Ti\emissiontype{I} &    1   \\
 5170 &   8.2 & & 5167 & Mg\emissiontype{I} &    2   \\
      &   8.2 & & 5167 & Fe\emissiontype{I} &   37   \\
      &   8.2 & & 5173 & Mg\emissiontype{I} &    2   \\
 5184 &   7.8 & & 5184 & Mg\emissiontype{I} &    2   \\
 5208 &   7.4 & & 5205 & Cr\emissiontype{I} &    7   \\
      &   7.4 & & 5205 & Fe\emissiontype{I} &    1   \\
      &   7.4 & & 5206 & Cr\emissiontype{I} &    7   \\
      &   7.4 & & 5208 & Cr\emissiontype{I} &    7   \\
 5246 &   4.6 & &      &  ?                 &   --   \\
 5268 &  10.2 & & 5270 & Fe\emissiontype{I} &   15   \\
 5296 &  11.3 & & 5293 & Cu\emissiontype{I} &    1   \\
 5326 &  13.8 & & 5328 & Fe\emissiontype{I} &   15   \\
      &  13.8 & & 5331 & O\emissiontype{I}  &   12   \\
 5344 &  12.2 & &      &  ?                 &   --   \\
 5370 &  11.1 & & 5371 & Fe\emissiontype{I} &   15   \\
 5378 &   3.5 & & 5383 & Fe\emissiontype{I} &  1146  \\
 5394 &   8.5 & & 5393 & Fe\emissiontype{I} &   553  \\
 5397 &   4.0 & & 5397 & Fe\emissiontype{I} &    15  \\
 5407 &  12.6 & & 5404 & Fe\emissiontype{I} &  1165  \\
      &  12.6 & & 5406 & Fe\emissiontype{I} &    15  \\
 5429 &  10.0 & & 5430 & Fe\emissiontype{I} &    15  \\
 5448 &   7.6 & & 5447 & Fe\emissiontype{I} &    15  \\
 5460 &   8.9 & & 5456 & Fe\emissiontype{I} &    15  \\
      &   8.9 & & 5461 & Hg\emissiontype{I} &     1  \\
 5488 &   3.6 & &      &  ?                 &    --  \\
 5510 &   6.7 & & 5506 & Mo\emissiontype{I} &     1  \\
 5534 &  15.5 & & 5533 & Mo\emissiontype{I} &     1  \\
 5702 &  32.5 & & 5700 & Cu\emissiontype{I} &     1  \\
 5784 &  39.8 & & 5782 & Cu\emissiontype{I} &     1  \\
 5856 &  21.0 & &      &  ?                 &    --  \\
 5870 &   4.9 & &      &  ?                 &    --  \\
 5896 &  40.2 & & 5890 & Na\emissiontype{I} &     1  \\
      &  40.2 & & 5896 & Na\emissiontype{I} &     1  \\
 6104 &   2.2 & & 6104 & Li\emissiontype{I} &     2  \\
 6152 &  10.0 & & 6156 & O\emissiontype{I}  &    10  \\
 6702 &   4.7 & & 6708 & Li\emissiontype{I} &     1  \\
 5500-6100 &  & &      & FeO                &   --   \\
\hline
\endlastfoot
\end{longtable}

\end{document}